\documentclass[preprint,superscriptaddress,aps]{revtex4}

\usepackage{amsfonts}
\usepackage{graphics}
\usepackage{graphicx}
\usepackage{slashed}
\usepackage{indentfirst}            
\usepackage{microtype}              
\usepackage{subeqnarray}            
\usepackage{amsmath}                
\usepackage{mathtools}
\usepackage{calrsfs}
\begin{document}
	
\newcommand{\be}{\begin{equation}}
\newcommand{\ee}{\end{equation}}
\newcommand{\en}{\end{equation}}
\newcommand{\ba}{\begin{eqnarray}}
\newcommand{\ea}{\end{eqnarray}}
\newcommand{\bea}{\begin{eqnarray}}
\newcommand{\eea}{\end{eqnarray}}
\newcommand{\bq}{\begin{eqnarray}}
\newcommand{\eq}{\end{eqnarray}}
\newcommand{\dt}{\frac{d^3k}{(2 \pi)^3}}
\newcommand{\dtp}{\frac{d^3p}{(2 \pi)^3}}
\newcommand{\kbruto}{\hbox{$k \!\!\!{\slash}$}}
\newcommand{\pbruto}{\hbox{$p \!\!\!{\slash}$}}
\newcommand{\qbruto}{\hbox{$q \!\!\!{\slash}$}}
\newcommand{\lbruto}{\hbox{$l \!\!\!{\slash}$}}
\newcommand{\bbruto}{\hbox{$b \!\!\!{\slash}$}}
\newcommand{\parbruto}{\hbox{$\partial \!\!\!{\slash}$}}
\newcommand{\Abruto}{\hbox{$A \!\!\!{\slash}$}}
\newcommand{\pa}{\partial}
\def\pls{\partial\!\!\!/}
\def\bb{\bibitem}
\def\as{a\!\!\!/}
\def\As{A\!\!\!/}
\def\ks{k\!\!\!/}
\def\ls{l\!\!\!/}
\def\ps{p\!\!\!/}
\def\qs{q\!\!\!/}
\def\bs{b\!\!\!/}
\def\yb{\bar{\y}}
\def\Ds{D\!\!\!\!/}
\def\ds{\partial\!\!\!/}
\newcommand{\fr}{\frac}
\def\ov{\over}
\def\g{\gamma}
\def\n{\nu}
\def\m{\mu}
\def\n{\nu}
\def\bb{\bibitem}
\def\eff{\mathrm{eff}}
\def\CS{\mathrm{CS}}
\newcommand{\Slash}[1]{{#1}\!\!/}
\newcommand{\SLASH}[1]{{#1}\!\!\!/}
\newcommand{\RM}[1]{\mathrm{#1}}

\title{On the nonlinear electrodynamics in a Lorentz-breaking scenario}

\date{today}

\author{E. Neres J\'unior} \email[]{edson.neres@ufvjm.edu.br.br}
\affiliation{Centro Federal de Educa\c{c}\~ao Tecnol\'ogica - MG, Avenida Amazonas, 7675 - 30510-000 - Nova Gameleira - Belo Horizonte-MG, Brasil}
\affiliation{Instituto de Engenharia, Ci\^encia e Tecnologia, Universidade Federal dos Vales do Jequitinhonha e Mucuri, Avenida Um, 4050 - 39447-790 -Cidade Universit\'aria -Jana\'uba-MG, Brasil}

\author{J. C. C. Felipe} \email[]{jeanccfelipe@ufsj.edu.br}
\affiliation{Instituto de Engenharia, Ci\^encia e Tecnologia, Universidade Federal dos Vales do Jequitinhonha e Mucuri, Avenida Um, 4050 - 39447-790 -Cidade Universit\'aria -Jana\'uba-MG, Brasil}
\affiliation{Departamento de Estat\' istica, F\'isica e Matem\'atica, Universidade Federal de São João del Rei, Rod. MG 443, Km 7, 36497-899 - Ouro Branco-MG, Brasil}

\author{A. P. Ba\^eta Scarpelli} \email[]{scarpelli@cefetmg.br}
\affiliation{Centro Federal de Educa\c{c}\~ao Tecnol\'ogica - MG, Avenida Amazonas, 7675 - 30510-000 - Nova Gameleira - Belo Horizonte-MG, Brasil}

\begin{abstract}
In this work, we study a model in nonlinear electrodynamics in the presence of a CPT-even term that violates Lorentz symmetry. The Lorentz-breaking vector, in addition to the usual background magnetic field, produces interesting effects in the dispersion relations. The consequences on the vacuum refractive index and the group velocity are studied. Vacuum birefringence is discussed in the case the nonlinear electrodynamics is a Euler-Heisenberg model.
\end{abstract}

\maketitle

\section{Introduction}

Nonlinear electrodynamics was first consistently investigated in the thirty's, when it was verified that vacuum effects, like the creation and annihilation of pairs, induce photon self-interactions. It was Born and Infeld \cite{Born2}, however, that elaborated the first generalization of Maxwell electrodynamics to a nonlinear version, with the aim of ensure that the self-energy of the electric field of charged particles is finite. It is possible to connect the origin of Born and Infeld's electrodynamics to the studies of Gustave Mie in 1909, who tried to create a purely electromagnetic theory of charged particles. The question of why the charged particle does not explode under the action of the repulsive Coulomb force, however, was raised even before \cite{Nora}. Born and Infeld's theory many years later became an important ingredient in string theory, quantum-gravitational models and theories with magnetic monopoles \cite{FRA,BER,BAN,GUN,BEATO,NIA,NIA2}.

The usual electrodynamics Lagrangian density is quadratic in the field strength tensor $F^{\mu\nu}$, so that the Maxwell equations are linear. Actions with higher order corrections in $F^{\mu\nu}$ will originate a nonlinear theory. In the Euler-Heisenberg model \cite{EH}, the higher order corrections are originated from the effect of one-loop calculations of the vacuum polarization. Actually, it is a full non-perturbative one-loop effective action for spinorial electrodynamics (see, for example, \cite{DMI}).

Nonlinear extensions of electrodynamics nowadays are used in the study of gravity, cosmology and condensed matter physics \cite{Hendi,Zhao,Miskovic,Habib,Aiello,Balart}. The renewed interest in nonlinear electrodynamics (see \cite{Hela1} and references therein) is also justified by the new very high intensity LASERs. Such devices are the best way to investigate classical and quantum electrodynamics in the strong field regime. The recent LASERs may, in the near future, be used in a detailed inspection of the structure of the vacuum and, therefore, of the electromagnetic nonlinearities \cite{Karbs,Markl,Battesti,Kink}.

In connection with the usual treatment of nonlinear electrodynamics, the study of Lorentz-violating models is also involved with questions like the anisotropy of the spacetime. This is because the Standard Model Extension (SME) \cite{SME1, SME2}, the model which is obtained by adding all possible Lorentz-violating terms which could emerge due to spontaneous breaking of this symmetry, establishes a set of constant tensors as the parameters of Lorentz violation which select special directions in spacetime. Recently, it was pursued the investigation of a generic nonlinear extension of axionic electrodynamics in a Carroll-Field-Jackiw (CFJ) \cite{CFJ} scenario that implements Lorentz-symmetry violation \cite{Hela2}. 

Considering the experimental aspects of Lorentz violation, there are imprints in quantum gravity at low energies \cite{Kost1,Dopli,Boj,Amel}. Experiments based on multi-messenger astronomy, for instance, bring us new possibilities to test diﬀerent models that incorporate Lorentz violation \cite{Adda}. In particular, from the perspective of the Standard Model Extension (SME) program, proposed tests involve gravitational waves,\cite{Kost2,Kost3,Schr} cosmic rays, \cite{Kost4,Alts} and astrophysical tests with neutrinos and gamma-ray photons \cite{Bert}. Fortunately, there is an extensive compilation of constraints on the values of the coeﬃcients for Lorentz and CPT violation in the SME \cite{Kost5,Kost-CPT}, permitting quantifying Lorentz violation from the diﬀerent sectors of the SME. In this paper, we study nonlinear electrodynamics under the influence of a specific kind of Lorentz-violating term. This is a special case of the CPT-even term of the photon sector in the SME.

The interest in the CPT-even terms has increased some time ago in connection with the aether concept and to the study of extra dimensions. It was discussed, for example \cite{Carroll}, the use of Lorentz-violating tensor fields with expectation values aligned with the extra dimensions so as to keep these extra
dimensions hidden. The Lorentz-violating terms of the gauge, fermion and scalar sectors, in this case, would come from the interaction between these fields with the aether fields. The kind of CPT-even term we are considering can be radiatively generated when a Lorentz-violating nonminimal coupling is introduced in electrodynamics \cite{Petrov,Scarp,Scarp2,Petrov-Scarp}.

In our study, we investigate how this CPT-even Lorentz-breaking term interferes with an external uniform and constant magnetic field. Dispersion relations are derived and the modified refraction indices and group velocities are calculated. Features like birefringence are discussed. Besides, we carry out an analysis of the influence of the background vectors in the wave propagation.

The paper is organized as follows: in section 2, we present the model and derive, up to second order in the photon field, the most general Lagrangian density. The field equations are obtained. In section 3 the dispersion relations are determined under special conditions. Then, the refractive indices and group velocities are obtained. In section 4, we discuss vacuum birefringence under the influence of the external fields for a Euler-Heisenberg model. Section 5 is left for the concluding comments.

\section{The nonlinear electrodynamics with a CPT-even Lorentz-breaking term}

Let us consider a model of nonlinear electrodynamics including a Lorentz-breaking term, described by the following Lagrangian density:
\begin{equation}
    {\cal L}= {\cal L}_{nl}({\cal F}_0,{\cal G}_0) - \frac 12( c_\mu F^{\mu\nu}_0)^2 - J_\mu A_0^\mu,
\end{equation}
in which ${\cal L}_{nl}({\cal F}_0,{\cal G}_0)$ is the most general Lagrangian for a nonlinear electrodynamics that depends on the Lorentz and gauge invariant bilinears, ${\cal F}_0=-\frac 14 F_{0\mu\nu}F_0^{\mu\nu}$ and ${\cal G}_0=-\frac 14 F_{0\mu\nu} \tilde{F}_0^{\mu\nu}$. In these definitions, $F_{0\mu\nu}=\partial_\mu A_{0\nu} - \partial_\nu A_{0\mu}$ is the field strength tensor and $\tilde{F}_0^{\mu\nu}=\frac 12 \epsilon^{\mu\nu\alpha\beta}F_{0\alpha\beta}$ is its correspondent dual tensor. Besides, the Lorentz-breaking term introduces the constant 4-vector $c^\mu$, which can be considered a background that selects a special direction in the spacetime.

In order to carry out our study, we expand the gauge field as $A_0^\mu = a^\mu + A_B^\mu$, splitting the electromagnetic field in the photon part, $a^\mu$, and the background potential $A_B^\mu$. At this stage, the background field is considered dependent on the spacetime coordinates. The expansion around the background field is carried out up to second order in the propagating field $a^\mu$, such that we obtain
\bq
{\cal L}^{(2)}= -\frac 14 \chi_1 f_{\mu\nu}^2 - \frac 14 \chi_2 f_{\mu\nu} \tilde{f}^{\mu\nu} +\frac 18 Q_{B\mu\nu\kappa\lambda}f^{\mu\nu}f^{\kappa \lambda} - \frac 12 (c_\mu f^{\mu\nu})^2 - \bar{J}_\mu a^\mu,
\eq
where
\be
\bar{J}_\mu = J_\mu -\partial^\nu G_{B\nu\mu} - \partial^\nu(c^\alpha c_\nu F_{B\alpha\mu} - c^\alpha c_\mu F_{B\alpha\nu})
\ee
is an effective external current which couples to the photon and the background tensors are defined as
\bq
G_{B\mu\nu}&=&\chi_1 F_{B\mu\nu} + \chi_2 \tilde{F}_{B\mu\nu} 
\eq
and
\bq
Q_{B\mu\nu\kappa\lambda}&=&d_1F_{B\mu\nu}F_{B\kappa\lambda} + d_2\tilde{F}_{B\mu\nu}\tilde{F}_{B\kappa\lambda} + d_3 F_{B\mu\nu}\tilde{F}_{B\kappa\lambda} + d_3\tilde{F}_{B\mu\nu}F_{B\kappa\lambda}.
\label{tensor}
\eq
The coefficients of the expansion are calculated in the background fields, $\boldsymbol{E}$ and $\boldsymbol{B}$, and are given by
\bq
\chi_1 &=& \left. \frac{\partial {\cal L}_{nl}}{\partial {\cal F}_0} \right|_{\boldsymbol{E},\boldsymbol{B}}, \,\,\,\,
\chi_2 = \left. \frac{\partial {\cal L}_{nl}}{\partial {\cal G}_0} \right|_{\boldsymbol{E},\boldsymbol{B}}, \,\,\,\,
d_1 = \left. \frac{\partial^2 {\cal L}_{nl}}{\partial {\cal F}_0^2} \right|_{\boldsymbol{E},\boldsymbol{B}}, \nonumber \\ 
d_2&=& \left. \frac{\partial^2 {\cal L}_{nl}}{\partial {\cal G}_0^2} \right|_{\boldsymbol{E},\boldsymbol{B}}, \,\,\,\,
d_3= \left. \frac{\partial^2 {\cal L}_{nl}}{\partial {\cal F}_0\partial {\cal G}_0} \right|_{\boldsymbol{E},\boldsymbol{B}},
\eq
which can be functions of the spacetime coordinates. It is important to note that we did not considered, in the Lagrangian density ${\cal L}^{(2)}$, the terms which depend only on the background fields, since they are irrelevant for the field equations we are interested in. 

A relevant discussion is concerned with the similarity between the terms in $Q_{B\mu\nu\kappa\lambda}f^{\mu\nu}f^{\kappa \lambda}$ and $(c_\mu f^{\mu\nu})^2$. In fact, we can define a tensor
\be
\kappa_{\mu\nu\alpha\beta}= \eta_{\mu\alpha}c_\nu c_\beta - \eta_{\nu\alpha} c_\mu c_\beta + \eta_{\nu\beta}c_\mu c_\alpha - \eta_{\mu\beta}c_\nu c_\alpha,
\ee
such that $(c_\mu f^{\mu\nu})^2 \sim \kappa_{\mu\nu\alpha\beta}f^{\mu\nu}f^{\alpha \beta}$. So, the non linearity induced by external electromagnetic fields is expected to produce similar effects as the ones which would be caused by some kinds of Lorentz-violation. Therefore, it is interesting to examine how these two backgrounds interfere with each other.

The similarity between the Lorentz-violating term and the one which involves the tensor defined in equation \eqref{tensor} raises a new discussion. Why not redefine the tensor $Q_{B\mu\nu\kappa\lambda}$ in such a way to absorb the dependence on $c^\mu$? There are famous papers like \cite{Kost-Mewes-1}, \cite{Kost-Mewes-2} and \cite{Kost-Mewes-3} in which a deep analysis is performed for a model like that, departing from general dispertion relations. First of all, it is important to note that the nature of the two terms is completely diverse. While the Lorentz-symmetry violation term would be originated from a more fundamental model and would be a characteristic of spacetime, the $Q_B$ tensor depends on the background magnetic field, which is a controlled parameter. Besides, the results obtained in \cite{Kost-Mewes-1}, for example, are at the leading order in the tensor $\kappa_{\mu\nu\alpha\beta}$, which is totally adequate in that case, since the magnitude of the the Lorentz-violating parameter is severely constrained by experimental results. However, in the present case, we may be dealing with very strong magnetic fields, for which such approximation is not possible.

We now derive the field equations, by using the action principle in relation to $a^\mu$, which reads
\be
\partial^\mu \left[\chi_1 f_{\mu\nu} + \chi_2 \tilde{f}_{\mu\nu} -\frac 12 Q_{B\mu\nu\kappa\lambda}f^{\kappa\lambda} + c^\alpha(c_\mu f_{\alpha\nu} - c_\nu f_{\alpha\mu}) \right] = \bar{J}_\nu.
\label{fequations}
\ee
Besides, we have that the Bianchi identity, $\partial_\mu \tilde{f}^{\mu\nu}=0$, remains valid. In the next section, we will carry out the calculation of the dispersion relations for particular situations of interest. It will be investigated the situation in which the background electromagnetic field reduces to an uniform magnetic field $\boldsymbol{B}$.

It is noteworth to observe that the coefficient $\chi_2$ vanishes in models which preserve CPT invariance, like the one under investigation. Furthermore, in all the studied nonlinear models for the electrodynamics, the coefficiente $d_3$ is null. We then will not take in consideration these coefficients from now on.

\section{Dispertion relations}

In this section we will obtain the dispersion relations for particular cases of interest in the situation where we have a background uniform magnetic field in the absence of sources ($\mathbf{\bar{J}}=\mathbf{0}$ and $\bar{\rho}=0$). Besides, we take a purely spacelike Lorentz-violating vector, $c^\mu=(0,\boldsymbol{c})$. 

We first derive the modified Gauss equation by taking $\nu=0$ in equation (\ref{fequations}). We obtain
\be
\boldsymbol{\nabla}\cdot \boldsymbol{D}=0, \label{Gauss}
\ee
with 
\be
\boldsymbol{D}=\boldsymbol{e}+ \frac{d_2}{\chi_1}(\boldsymbol{B}\cdot \boldsymbol{e})\boldsymbol{B} - \frac{1}{\chi_1} (\boldsymbol{c}\cdot \boldsymbol{e}) \boldsymbol{c}.
\ee
For the Amp\`ere-Maxwell equation, we take the spatial index $i$ in equation (\ref{fequations}), so that
\be
\boldsymbol{\nabla} \times \boldsymbol{H} - \frac{1}{\chi_1}(\boldsymbol{c} \cdot \boldsymbol{\nabla})(\boldsymbol{c}\times \boldsymbol{b}) + \frac{1}{\chi_1}\boldsymbol{\nabla}\cdot(\boldsymbol{c}\times \boldsymbol{b})\boldsymbol{c}= \frac{\partial \boldsymbol{D}}{\partial t}, \label{Ampere}
\ee
where
\be
\boldsymbol{H}=\boldsymbol{b} - \frac{d_1}{\chi_1} (\boldsymbol{b}\cdot \boldsymbol{B}) \boldsymbol{B}.
\ee
From the Bianchi identity, the two usual homogeneous equations are written as
\be
\boldsymbol{\nabla}\times \boldsymbol{e} = - \frac{\partial \boldsymbol{b}}{\partial t}
\;\;\;\; \mbox{and} \;\;\;\; \boldsymbol{\nabla} \cdot \boldsymbol{b} = 0.
\ee

Next, with the aim of obtaining the dispersion relations, we eliminate the magnetic field $\boldsymbol{b}$ from the equations. For this, we take the time derivative of Amp\`ere-Maxwell equation (\ref{Ampere}) and use Faraday-Lenz relation to obtain an equation for $\boldsymbol{e}$. We then try a plane wave solution $\boldsymbol{e}(\boldsymbol{x},t)=\boldsymbol{e}_0 e^{i(\boldsymbol{k}\cdot \boldsymbol{x}- \omega t)}$ and obtain
\be
M_{ij} e_{0j}=0,
\label{wave}
\ee
with
\bq
M_{ij}&=& \left[\omega^2 - \boldsymbol{k}^2 + \frac{1}{\chi_1} (\boldsymbol{c}\cdot \boldsymbol{k})^2 \right]\delta_{ij} +\frac{d_1}{\chi_1} (\boldsymbol{B}\times \boldsymbol{k})_i (\boldsymbol{B}\times \boldsymbol{k})_j -\frac{1}{\chi_1}(\boldsymbol{c}\cdot \boldsymbol{k})c_i k_j + \nonumber \\
&-& \frac{1}{c_1}(\omega^2 - \boldsymbol{k}^2)c_i c_j - \frac{d_2}{\chi_1} (\boldsymbol{B}\cdot \boldsymbol{k})k_i B_j + \frac{d_2}{\chi_1} \omega^2 B_i B_j.
\label{mmatrix}
\eq

The dispersion relations can be found by imposing a nontrivial solution for equation (\ref{wave}). This is accomplished with the requirement that $\det M=0$, which gives us the following equation:
\bq
&& (\omega^2 - \boldsymbol{k}^2) \left\{ (\omega^2 - \lambda \boldsymbol{k}^2) \left\{(\omega^2 - \lambda \boldsymbol{k}^2)f +\frac{d_2}{\chi_1}\left[\omega^2\boldsymbol{B}^2 -\boldsymbol{k}^2(\boldsymbol{B}\cdot \hat{\boldsymbol{k}})^2\right]f +\frac{d_1}{\chi_1}\boldsymbol{k}^2(\boldsymbol{B}\times \hat{\boldsymbol{k}})^2f + \right. \right.\nonumber \\ 
&& + \left. \left. \frac{d_1}{\chi_1^2}\boldsymbol{k}^2(\boldsymbol{B}\times \hat{\boldsymbol{k}}\cdot \boldsymbol{c})^2
+  \frac{d_2}{\chi_1^2}(\boldsymbol{B}\cdot \boldsymbol{c})^2 \omega^2 \right\}
+ \frac{d_1 d_2}{\chi_1^2}\boldsymbol{k}^2(\boldsymbol{B}\times \hat{\boldsymbol{k}})^2 \left[\omega^2\boldsymbol{B}^2 -\boldsymbol{k}^2(\boldsymbol{B}\cdot \hat{\boldsymbol{k}})^2\right]f +
\right.  \nonumber \\
&& \left.+ \frac{d_1 d_2}{\chi_1^3} \left[ \omega^2(\boldsymbol{B}\cdot \boldsymbol{c})^2 (\boldsymbol{B}\times \hat{\boldsymbol{k}})^2
+ \boldsymbol{k}^2(\boldsymbol{B}\times \hat{\boldsymbol{k}}\cdot \boldsymbol{c})^2 \left[\omega^2\boldsymbol{B}^2 -\boldsymbol{k}^2(\boldsymbol{B}\cdot \hat{\boldsymbol{k}})^2\right]
\right]\right\}=0,
\eq
where $f= 1 - \frac{\boldsymbol{c}^2}{\chi_1}$, $\lambda = 1 - \frac{(\boldsymbol{c} \cdot \hat{\boldsymbol{k}})^2}{\chi_1}$ and $\hat{\boldsymbol{k}}= \boldsymbol{k}/|\boldsymbol{k}|$. The first solution is the conventional frequency of Lorentz invariant linear electrodynamics, $\omega_1 =|\boldsymbol{k}|$. For the another factor, we write the equation, in order to obtain a general solution \cite{Visser}, as
\be
P \omega^4+ Q \omega^2 + R = 0,
\ee
with
\bq
P &=& fH + \frac{d_2}{\chi_1}(\boldsymbol{B}\cdot \boldsymbol{c})^2, \nonumber \\
Q &=&  \boldsymbol{k}^2 \left[ AH - \frac{d_2}{\chi_1}(\boldsymbol{B}\cdot \boldsymbol{c})^2 F - fG
\right] \;\;\;\;\; \mbox{and} \nonumber \\
R &=& - GA \boldsymbol{k}^4,
\eq
where
\bq
F &=& 1 - \frac{1}{\chi_1}(\boldsymbol{c}\cdot \hat{\boldsymbol{k}})^2 - \frac{d_1}{\chi_1}(\boldsymbol{B}\times \hat{\boldsymbol{k}})^2 ,\;\;\;\;
G = 1 - \frac{1}{\chi_1}(\boldsymbol{c}\cdot \hat{\boldsymbol{k}})^2 + \frac{d_2}{\chi_1}(\boldsymbol{B}\cdot \hat{\boldsymbol{k}})^2, \nonumber \\
A &=& -fF + \frac{d_1}{\chi_1^2}(\boldsymbol{B}\times \hat{\boldsymbol{k}} \cdot \boldsymbol{c})^2 \;\;\;\; \mbox{and} \;\;\;\; H = 1 + \frac{d_2}{\chi_1}\boldsymbol{B}^2.
\eq
The two positive solutions are given by
\be
\omega_{2,3}= |\boldsymbol{k}|\left\{\frac{1}{2\left[fH+\frac{d_2}{\chi_1}(\boldsymbol{B}\cdot \boldsymbol{c})^{2}\right]}\left\{-AH+fG+\frac{d_{2}}{\chi_{1}^{2}}(\boldsymbol{B}\cdot \boldsymbol{c})^{2}F \pm \sqrt{\tilde{\Delta}}\right\}\right\}^{1/2},
\ee
with
\be
\tilde{\Delta}= \left[ AH + fG- \frac{d_2}{\chi_1}(\boldsymbol{B}\cdot \boldsymbol{c})^2 F \right]^2 +
4 \frac{d_1 d_2}{\chi_1^3}(\boldsymbol{B}\cdot \boldsymbol{c})^{2}(\boldsymbol{B}\times \hat{\boldsymbol{k}} \cdot \boldsymbol{c})^2 G.
\ee

The expressions above for the two last frequencies are complicated and of difficult interpretation. We, then, study below some particular situations.

\subsection{The case in which the plane of $\boldsymbol{B}$ and $\boldsymbol{k}$ is orthogonal to the Lorentz-violating vector}

In this case, the geometry of the problem gives us $\boldsymbol{c} \cdot \boldsymbol{B}=\boldsymbol{c} \cdot \boldsymbol{k}=0$ and the frequencies are reduced to the simple forms below:
\bq
\omega_1 &=& |\boldsymbol{k}|, \\
\omega_2 &=& |\boldsymbol{k}| \sqrt{1 - \frac{d_2(\boldsymbol{B}\times \boldsymbol{\hat{k}}) ^2}{\chi_1 + d_2\boldsymbol{B}^2}}, \\
\omega_3 &=& |\boldsymbol{k}|\sqrt{1 - \frac{d_1}{f \chi_1}(\boldsymbol{B}\times \boldsymbol{\hat{k}}) ^2}.
\eq
The first frequency is the conventional one of Lorentz invariant linear electrodynamics, as already observed in the general case. Furthermore, the two other frequencies are exactly the same calculated in \cite{Hela1} in which only the uniform and constant background magnetic field was considered. Only the third frequency is affected by the Lorentz breaking vector $\boldsymbol{c}$. In this case the Lorentz invariant result is easily recovered by making $\boldsymbol{c}=\boldsymbol{0}$.

We now determine the refraction indices. The vacuum refraction index is given by the inverse of the phase velocity, $\omega_i/|\boldsymbol{k}|$. We then have for the two new frequencies
\bq
n_2^{-1} &=& \sqrt{1 - \frac{d_2(\boldsymbol{B}\times \boldsymbol{\hat{k}}) ^2}{\chi_1 + d_2\boldsymbol{B}^2}},\\
n_3^{-1}&=&  \sqrt{1 - \frac{d_1}{f \chi_1}(\boldsymbol{B}\times \boldsymbol{\hat{k}}) ^2}.
\eq

For the group velocities, we have $v_{gi}=\frac{\partial w}{\partial k_i}$. The results read
\bq
\left.\boldsymbol{v}_{g}\right|_{\omega=\omega_2} &=& \frac{\chi_1 \boldsymbol{\hat{k}} + d_2 (\boldsymbol{B}\cdot \boldsymbol{\hat{k}})\boldsymbol{B}}{(\chi_1+d_2 \boldsymbol{B}^2) \sqrt{1- \frac{d_2(\boldsymbol{B} \times \boldsymbol{\hat{k}})^2}{\chi_1+d_2 \boldsymbol{B}^2}}}, \\
\left.\boldsymbol{v}_{g}\right|_{\omega=\omega_3} &=& \frac{\chi_1 f \boldsymbol{\hat{k}} + d_1 \boldsymbol{B}\times(\boldsymbol{B}\times \boldsymbol{\hat{k}})}{\chi_1 f \sqrt{1-\frac{d_1}{\chi_1 f}(\boldsymbol{B} \times \boldsymbol{\hat{k}})^2}}.
\eq
Again, it is very direct to recover the results of literature in the limits of Lorentz invariance and of linear electrodynamics.

\subsection{The case in which $\boldsymbol{B}$, $\boldsymbol{k}$ and the Lorentz-violating vector are coplanar}

Next, we study the more involved situation in which the the three vectors, $\boldsymbol{c}$, $\boldsymbol{B}$ and $\boldsymbol{k}$, are coplanar. We first obtain the frequencies for the most general configuration. For this case the mixed product $\boldsymbol{B}\times \hat{\boldsymbol{k}} \cdot \boldsymbol{c}$ is null, so that $\tilde{\Delta}$ is a perfect square. The frequencies reduce to
\bq
\omega_1 &=& |\boldsymbol{k}|, \nonumber \\
\omega_2 &=& |\boldsymbol{k}|\left\{\frac{f}{fH+\frac{d_2}{\chi_1}(\boldsymbol{B}\cdot \boldsymbol{c})^{2}}\left[1 - \frac{1}{\chi_1}(\boldsymbol{c}\cdot \hat{\boldsymbol{k}})^2 + \frac{d_2}{\chi_1}(\boldsymbol{B}\cdot \hat{\boldsymbol{k}})^2\right]\right\}^{1/2}, \nonumber \\
\omega_3 &=& |\boldsymbol{k}|\left\{ 1 - \frac{1}{\chi_1}(\boldsymbol{c}\cdot \hat{\boldsymbol{k}})^2 - \frac{d_1}{\chi_1}(\boldsymbol{B}\times \hat{\boldsymbol{k}})^2\right\}^{1/2}.
\eq

The refractive indices and the group velocities for the non conventional frequencies are directly obtained as
\bq
n_2^{-1} &=& \left\{\frac{f}{fH+\frac{d_2}{\chi_1}(\boldsymbol{B}\cdot \boldsymbol{c})^{2}}\left[1 - \frac{1}{\chi_1}(\boldsymbol{c}\cdot \hat{\boldsymbol{k}})^2 + \frac{d_2}{\chi_1}(\boldsymbol{B}\cdot \hat{\boldsymbol{k}})^2\right]\right\}^{1/2}, \nonumber \\
n_3^{-1} &=& \left\{ 1 - \frac{1}{\chi_1}(\boldsymbol{c}\cdot \hat{\boldsymbol{k}})^2 - \frac{d_1}{\chi_1}(\boldsymbol{B}\times \hat{\boldsymbol{k}})^2\right\}^{1/2}
\eq
and
\bq
\left.\boldsymbol{v}_{g}\right|_{\omega=\omega_2} &=& 
\frac{n_2}{fH+\frac{d_2}{\chi_1}(\boldsymbol{B}\cdot \boldsymbol{c})^{2}} \left\{\hat{\boldsymbol{k}} + 
\frac{1}{\chi_1}\left[d_2(\boldsymbol{B}\cdot \hat{\boldsymbol{k}}) \boldsymbol{B} - (\boldsymbol{c}\cdot \hat{\boldsymbol{k}}) \boldsymbol{c}\right]
\right\}, \nonumber \\
\left.\boldsymbol{v}_{g}\right|_{\omega=\omega_3} &=& 
n_3 \left\{H\hat{\boldsymbol{k}} + 
\frac{1}{\chi_1}\left[d_2(\boldsymbol{B}\cdot \hat{\boldsymbol{k}}) \boldsymbol{B} - (\boldsymbol{c}\cdot \hat{\boldsymbol{k}}) \boldsymbol{c}\right]
\right\}.
\eq
Let us note that the velocity group is a linear combination of the three vectors, $\boldsymbol{c}$, $\boldsymbol{B}$ and $ \boldsymbol{k}$.

From the most general results for the coplanar problem presented above, we go, in the subsections below, to more specific situations which we think it is interesting to analyze.

\subsubsection{First case: $\boldsymbol{c}\bot \boldsymbol{B}$, $\boldsymbol{c}\bot \boldsymbol{k}$ and $\boldsymbol{B}\parallel \boldsymbol{k}$}

In this simple but interesting case, we obtain that
the unique possible solution is the conventional one $\omega^2=\boldsymbol{k}^2$. The interesting fact is that the two backgrounds (magnetic field and Lorentz-breaking vector), in this case, do not affect the solutions of frequency. Consequently, the refraction index and the group velocity are also preserved.

\subsubsection{Second case: $\boldsymbol{c} \parallel \boldsymbol{B} \parallel \boldsymbol{k}$}

This second case, at first glance, seems to be very simple. However, this geometrical configuration allows us to see how the Lorentz violation is decisive in the modification of wave propagation. The obtained frequencies are given below:
\bq
\omega_1(\boldsymbol{k}) &=& |\boldsymbol{k}| ,\\
\omega_2(\boldsymbol{k}) &=& \omega_3 (\boldsymbol{k})  = \sqrt{f}|\boldsymbol{k}|.
\eq

The first analysis to be done is about the limits in order to establish comparisons with the known results of the literature. If we restore Lorentz invariance by making $\boldsymbol{c}= \boldsymbol{0}$, we recover the results of Maxwell theory, even maintaining the background magnetic field. This is in accord with \cite{Hela1}, if we take $\boldsymbol{B}$ parallel to $\boldsymbol{k}$. On the other hand, we can also recover the situation in which we have a linear electrodynamics but with the breaking of Lorentz symmetry. First, it is important to note that this is accomplished by taking $\boldsymbol{B}\to \boldsymbol{0}$. In this situation, in all nonlinear known models, we have $\chi_1=1$, $d_1=d_2=0$. Taking these limits, the known results for the Lorentz-violating electrodynamics we are considering are recovered.

For the refraction indices, we get
\bq
n_1^{-1} &=& 1 ,\\
n_2^{-1}&=& n_3^{-1} = \sqrt{f}.
\eq
The group velocities are also directly obtained:
\bq
\left.\boldsymbol{v}_{g}\right|_{\omega=\omega_1} &=& \hat{\boldsymbol{k}}, \\
\left.\boldsymbol{v}_{g}\right|_{\omega=\omega_2} &=& \left.\boldsymbol{v}_{g}\right|_{\omega=\omega_3} = \hat{\boldsymbol{k}}\sqrt{f}.
\eq
Again, it is very direct to recover the results of literature in the limits of Lorentz invariance and of linear electrodynamics.

\subsubsection{Third case: $\boldsymbol{c}\parallel \boldsymbol{B}$ and $\boldsymbol{c}, \boldsymbol{B} \bot \boldsymbol{k}$}

In this third case, the external magnetic field is the one decisive for the modification of the wave propagation. Besides the usual solution $w^2=\boldsymbol{k}^2$, we obtain
\bq
\omega_2 &=& |\boldsymbol{k}|\sqrt{\frac{f}{f +\frac{d_2}{\chi_1}\boldsymbol{B}^2}}, \\
\omega_3 &=& |\boldsymbol{k}| \sqrt{1-\frac{d_1}{\chi_1}\boldsymbol{B}^2},
\eq
 For a null background magnetic field, the results of conventional electrodynamics are recovered. It is noteworth that a Lorentz-violating vector orthogonal to the wave vector only affects the wave propagation if the background magnetic field is present. This can also be noted in the case where $\boldsymbol{c}$ is orthogonal to the plane of $\boldsymbol{B}$ and $\boldsymbol{k}$. 

The refraction indices and group velocities are straightforward to obtain:
\bq
n_2^{-1} &=& \sqrt{\frac{f}{f +\frac{d_2}{\chi_1}\boldsymbol{B}^2}},\\
n_3^{-1}&=&   \sqrt{1-\frac{d_1}{\chi_1}\boldsymbol{B}^2},
\eq
\bq
\left.\boldsymbol{v}_{g}\right|_{\omega=\omega_2} &=& \hat{\boldsymbol{k}}\sqrt{\frac{f}{f +\frac{d_2}{\chi_1}\boldsymbol{B}^2}}, \\
\left.\boldsymbol{v}_{g}\right|_{\omega=\omega_3} &=& \hat{\boldsymbol{k}}   \sqrt{1-\frac{d_1}{\chi_1}\boldsymbol{B}^2}.
\eq

\section{Birefringence in the Euler-Heisenberg model}

Up to now, we treated a general nonlinear model. Let us now deal with the specific case of the Euler-Heisenberg Lagrangian, which is given by
\be
{\cal L}_{nl}= {\cal F} + \frac{2 \alpha^2}{45 m^4}(4 {\cal F}^2 + 7 {\cal G}^2),
\ee
in which $\alpha$ is the fine structure constant and $m$ is the electron mass. In this case, we get
\be
\chi_1 = 1 - \frac{8 \alpha^2 \boldsymbol{B}^2}{45 m^4}, 
\;\;\; d_1= \frac{16 \alpha^2}{45 m^4} \;\;\; \mbox{and} \;\;\;
d_2= \frac{28 \alpha^2}{45 m^4}.
\ee

Next, we study the birefringence effects induced by the nonlinearity and by the Lorentz symmetry breaking vector. Birefringence is characterized by a refractive index dependent on the propagation direction of the wave. As in the last section we considered several possibilities of geometric configurations, we now analyze birefringence in two situations. First, we take the Lorentz-breaking vector orthogonal to the plane of $\boldsymbol{B}$ and $\boldsymbol{k}$ and, then consider the difference in the refractive index when the wave vector is parallel and orthogonal to the background magnetic field. The second situation is when we take $\boldsymbol{c}$ and $\boldsymbol{B}$ parallel. The difference in the refractive index is investigated, as before, when the wave propagation is taken parallel and perpendicular to the magnetic field. 

\subsection{the plane of $\boldsymbol{B}$ and $\boldsymbol{k}$ is perpendicular to $\boldsymbol{c}$}

If $\boldsymbol{B} \bot \boldsymbol{k}$, for an uniform external magnetic field, we have for the refractive indices:
\bq
n_{1\bot} &=& 1, \\
n_{2\bot} &=& \sqrt{\frac{1 +\frac 49 \frac{\alpha^2 \boldsymbol{B}^2}{m^4}}{1 -\frac {8}{45} \frac{\alpha^2 \boldsymbol{B}^2}{m^4}}}, \\
n_{3\bot} &=& \sqrt{\frac{1 -\frac{8}{45} \frac{\alpha^2 \boldsymbol{B}^2}{m^4}-\boldsymbol{c}^2}{1 -\frac{8}{15} \frac{\alpha^2 \boldsymbol{B}^2}{m^4}-\boldsymbol{c}^2}}.
\eq
On the other hand, if $\boldsymbol{k}\parallel \boldsymbol{B}$, we have
\be
n_{1\parallel} = n_{2\parallel} = n_{3\parallel} = 1.
\ee
The birefringence manifests itself as the difference between the refractive indices. We then have a matrix with elements
\be
\Delta_{ij} = n_{i\bot} - n_{j\parallel},
\ee
such that for $\boldsymbol{B}$ and $\boldsymbol{k}$ orthogonal to the background Lorentz-violating vector, we obtain
\bq
\Delta_{11} &=& \Delta_{12} = \Delta_{13} = 0, \nonumber \\
\Delta_{21} &=& \Delta_{22} = \Delta_{23} = \sqrt{\frac{1 +\frac 49 \frac{\alpha^2 \boldsymbol{B}^2}{m^4}}{1 -\frac {8}{45} \frac{\alpha^2 \boldsymbol{B}^2}{m^4}}}-1, \nonumber \\
\Delta_{31} &=& \Delta_{32} = \Delta_{33} = \sqrt{\frac{1 -\frac{8}{45} \frac{\alpha^2 \boldsymbol{B}^2}{m^4}-\boldsymbol{c}^2}{1 -\frac{8}{15} \frac{\alpha^2 \boldsymbol{B}^2}{m^4}-\boldsymbol{c}^2}}-1.
\eq
We observe that the Lorentz violation only affect the $\Delta_{3j}$ elements.

\subsection{$\boldsymbol{B}$ and $\boldsymbol{k}$ are coplanar with $\boldsymbol{c}$}
Here we choose to take $\boldsymbol{c}$ parallel to $\boldsymbol{B}$ and then make both orthogonal and parallel to $\boldsymbol{k}$. In the first case, we have
\bq
n_{1\bot} &=& 1,\\
n_{2\bot} &=& \sqrt{\frac{1 +\frac 49 \frac{\alpha^2 \boldsymbol{B}^2}{m^4}- \boldsymbol{c}^2}{1 -\frac {8}{45} \frac{\alpha^2 \boldsymbol{B}^2}{m^4}- \boldsymbol{c}^2}}, \\
n_{3\bot} &=& \sqrt{\frac{1 -\frac {8}{45} \frac{\alpha^2 \boldsymbol{B}^2}{m^4}}{1 - \frac {8}{15} \frac{\alpha^2 \boldsymbol{B}^2}{m^4}}},
\eq

If we take $\boldsymbol{k}$ parallel to $\boldsymbol{B}$, we obtain
\bq
n_{1\parallel} &=& 1, \\
n_{2\parallel} = n_{3\parallel} &=& \sqrt{\frac{1 - \frac {8}{45} \frac{\alpha^2 \boldsymbol{B}^2}{m^4}}{1 - \frac {8}{45} \frac{\alpha^2 \boldsymbol{B}^2}{m^4} - \boldsymbol{c}^2}}.
\eq
The matrix elements of the difference of refractive indices are then given by
\bq
\Delta_{11}&=&0, \nonumber \\
\Delta_{12} &=& \Delta_{13} = 1 - \sqrt{\frac{1 - \frac {8}{45} \frac{\alpha^2 \boldsymbol{B}^2}{m^4}}{1 - \frac {8}{45} \frac{\alpha^2 \boldsymbol{B}^2}{m^4} - \boldsymbol{c}^2}}, \nonumber \\
\Delta_{21} &=& \sqrt{\frac{1 +\frac 49 \frac{\alpha^2 \boldsymbol{B}^2}{m^4}- \boldsymbol{c}^2}{1 -\frac {8}{45} \frac{\alpha^2 \boldsymbol{B}^2}{m^4}- \boldsymbol{c}^2}} -1, \nonumber \\
\Delta_{22} &=& \Delta_{23} = \sqrt{\frac{1 +\frac 49 \frac{\alpha^2 \boldsymbol{B}^2}{m^4}- \boldsymbol{c}^2}{1 -\frac {8}{45} \frac{\alpha^2 \boldsymbol{B}^2}{m^4}- \boldsymbol{c}^2}} - \sqrt{\frac{1 - \frac {8}{45} \frac{\alpha^2 \boldsymbol{B}^2}{m^4}}{1 - \frac {8}{45} \frac{\alpha^2 \boldsymbol{B}^2}{m^4} - \boldsymbol{c}^2}},\nonumber \\
\Delta_{31} &=& \sqrt{\frac{1 -\frac {8}{45} \frac{\alpha^2 \boldsymbol{B}^2}{m^4}}{1 - \frac {8}{15} \frac{\alpha^2 \boldsymbol{B}^2}{m^4}}} - 1, \nonumber \\
\Delta_{32} &=& \Delta_{33} = \sqrt{\frac{1 -\frac {8}{45} \frac{\alpha^2 \boldsymbol{B}^2}{m^4}}{1 - \frac {8}{15} \frac{\alpha^2 \boldsymbol{B}^2}{m^4}}} - 
 \sqrt{\frac{1 - \frac {8}{45} \frac{\alpha^2 \boldsymbol{B}^2}{m^4}}{1 - \frac {8}{45} \frac{\alpha^2 \boldsymbol{B}^2}{m^4} - \boldsymbol{c}^2}}.
\eq

It is interesting to note how the magnetic field and the Lorentz-violating vector interfere with each other. In all the situations, $\boldsymbol{c}^2$ should be compared with $\frac{\alpha^2 \boldsymbol{B}^2}{m^4}$. It is expected that $\boldsymbol{c}^2$ is a very small parameter. For a magnetic field of the order of $1 T \sim 10^{-15}\, GeV^2$, we have $\frac{\alpha^2 \boldsymbol{B}^2}{m^4} \sim 10^{-22}$, which may be compatible with the magnitude of the Lorentz-breaking vector and with the result of magnetic vacuum birefringence, of the order of $10^{-24}$ \cite{Hela2}. On the other hand, for magnetic fields already of the order of $B \sim 10^2 T \sim 10^{-13} GeV^2$, we have $\frac{\alpha^2 \boldsymbol{B}^2}{m^4} \sim 10^{-18}$, and the influence of the Lorentz violation begins to become negligible. 

In the case of birefringence of the vacuum, the dispersion of light produced by galactic and extragalactic objects provide bounds of the order of $10^{-16}$ on certain coefficients for Lorentz violation in the photon sector \cite{Kost-Mewes-2}. 
From the Data Tables \cite{Kost5}, we can see that the experimental bounds on the coeﬃcients of the SME are very restrictive. In particular, for the dimensionless coeﬃcients in which we are interested in this paper, we have that  $\kappa^{\mu\nu\lambda\rho}$ is bound to about $10^{-17}$.


Next, let us see how the birefringence is manifest if the magnetic field is absent. Since we do not have the $\boldsymbol{B}$ vector, we are left only with the relative direction of $\boldsymbol{k}$ in comparison with $\boldsymbol{c}$. We will compare the situations where the vectors are perpendicular and parallel. For $\boldsymbol{c}\bot \boldsymbol{k}$, there is no modification in the refractive indices, which are all $1$. However, when the wave propagates parallel to $\boldsymbol{c}$, there is a slight increase in the refractive index, since we obtain
\bq
n_{1\parallel} &=& 1, \\
n_{2\parallel} = n_{3\parallel} &=& \sqrt{\frac{1 }{1 - \boldsymbol{c}^2}}.
\eq

\section{Final considerations}

We studied a general nonlinear model for the electrodynamics with the addition of a Lorentz-breaking CPT-even term. This term is parametrized by the 4-vector $c^\mu$ which selects a special direction in the spacetime. We considered the case in which the external field is an uniform and constant magnetic field and where the Lorentz-violating vector is constant and purely space-like ($c^\mu=(0, \boldsymbol{c})$). We found that the two backgrounds, magnetic and Lorentz-violating, affect the wave propagation. We analyzed the cases in which the magnetic field, $\boldsymbol{B}$ and the wave vector $\boldsymbol{k}$ are orthogonal to $\boldsymbol{c}$ or coplanar with this Lorentz-breaking vector. In the case the three vectors are coplanar, three situations were considered: in the first one, the vector $\boldsymbol{c}$ is orthogonal to both $\boldsymbol{B}$ and $\boldsymbol{k}$, which are parallel between themselves; in the second case, the three vectors are parallel to each other; and, in the third case, $\boldsymbol{k}$ is orthogonal to both $\boldsymbol{B}$ and $\boldsymbol{c}$, which are parallel between themselves. In the case the wave propagates parallel to the magnetic field but orthogonal to the Lorentz-violating vector, the dispersion relations are surprisingly preserved. It was not the case in the other studied situations, in which it was verified modifications in the dispersion relations and, consequently, in the vacuum refraction indices and in the group velocities. 

In view of the effects of the external magnetic field and Lorentz violation on the wave propagation, discussions on birefringence were carried out for a Euler-Heisenberg model. This was done by comparing the refractive indices when the photon propagates parallel and orthogonal to the magnetic field. This, in turn, was done in two situations: $\boldsymbol{B}$ and $\boldsymbol{k}$ are orthogonal to $\boldsymbol{c}$; and $\boldsymbol{B}$ and $\boldsymbol{k}$ are coplanar with $\boldsymbol{c}$. In all the cases, we found that the two external vectors interfere through the magnitudes of $\boldsymbol{c}^2$ and $\frac{\alpha^2 \boldsymbol{B}^2}{m^4}$. The effects of the Lorentz violation turns out to be negligible in comparison to the ones of a external magnetic field near the critical one. We also verified that, in the absence of the magnetic field, the wave propagation is not affected when it occurs in a direction perpendicular to the background vector $\boldsymbol{c}$.  

There are still some points that deserve to be investigated and that are under study. First, it is important to check how a timelike Lorentz-breaking vector, $c^\mu=(c_0,\boldsymbol{0})$, modifies the model. Furthermore, some physical effects of interest for application in condensed matter physics need to be discussed. Such points will be included in a future work.

\end{document}